# The Anatomy of Value Creation: Input-Output Linkages, Policy Shifts, and Economic Impacts

Sourish Dutta

sourish.dutta@vips.edu

19 November 2025

**Abstract**

This paper describes the construction of high-resolution 7-digit product-level Supply-Use Tables (SUTs) and symmetric Input-Output Tables (IOTs) for the Indian economy from 2016-17 to 2022-23, primarily utilising microdata from the Annual Survey of Industries (ASI). It outlines the intricate methodology for generating comprehensive input and output flows, reconciling data from registered and unregistered manufacturing sectors using NPCMS-NIC concordance, and transforming SUTs into analytically robust IOTs via the Industry Technology Assumption. The paper demonstrates the application of this framework through an in-depth case study of India's mobile phone manufacturing sector (NPCMS 4722200), analysing its economic impact, particularly on Domestic Value Added (DVA) and employment. The findings reveal significant structural shifts, including output growth, import substitution, export surges, and evolving labour dynamics. These detailed tables represent a valuable advancement for analysing sectoral interdependencies, the efficacy of industrial policy, and enhancing India's National Accounts Statistics, resonating with the NSS's legacy of data-driven policymaking.

## Introduction

The National Sample Survey (NSS), whose 75th anniversary we commemorate, stands as a monumental pillar in India's statistical system. Its rich legacy of providing robust, granular data has been instrumental in shaping our understanding of the Indian economy and in guiding evidence-based policymaking for decades. In this spirit of enhancing our analytical capabilities through detailed statistical frameworks, this paper focuses on the construction and application of highly disaggregated Supply-Use Tables (SUTs) and symmetric Input-Output Tables (IOTs).

The System of National Accounts (SNA), particularly the forthcoming SNA 2025, continues to emphasise the centrality of SUTs and IOTs (often detailed in dedicated chapters, e.g., Chapters 15 & 36 in SNA 2025 discussions) for comprehensive economic analysis and policymaking, especially in an increasingly interconnected global economy characterised by complex global value chains (GVCs). These tables are indispensable for dissecting economic structures, tracing inter-sectoral flows of goods and services, and evaluating the multifaceted impacts of economic policies or external shocks. They offer a coherent framework for reconciling data from diverse sources, thereby improving the consistency and accuracy of National Accounts Statistics (NAS). Moreover, their analytical dimension is significantly enhanced when SUTs are transformed into symmetric IOTs.

Despite their acknowledged importance, the availability of highly granular and regularly updated SUTs and IOTs remains a challenge in many developing countries, including India. While India has a long tradition of compiling IOTs, the level of disaggregation and frequency can continuously be improved to meet the evolving demands of policymakers and researchers. This paper contributes to addressing this gap by detailing the construction of 7-digit product-level SUTs and symmetric IOTs for the Indian economy covering the period 2016-17 to 2022-23, primarily using microdata from the Annual Survey of Industries (ASI). This level of granularity—distinguishing approximately 5,725 products —offers an unprecedented depth of detail for the Indian context, allowing for more nuanced analyses than previously feasible with more aggregated tables.

The core methodological contributions of this paper lie in (i) outlining the systematic generation of input flow matrices (Use tables) and output matrices (Make tables) by harmonising ASI data with other relevant datasets like KLEMS (which incorporates NSSO data for the unorganised sector) through meticulous concordance between product (NPCMS) and activity (NIC) classifications; (ii) explaining the mathematical transformation of these SUTs into symmetric product-by-product IOTs using the Industry Technology Assumption (ITA); and (iii) demonstrating the analytical utility of these tables by estimating Domestic Value Added (DVA), Foreign Value Added (FVA), and employment linkages (direct and indirect) arising from production and exports.

To illustrate the power of this framework, we present an in-depth case study of India's mobile phone manufacturing sector (NPCMS 4722200) for the period from 2016-17 to 2022-23. This sector has been a focal point of India's industrial policy, encompassing initiatives such as "Make in India" and the Production Linked Incentive (PLI) schemes. Our analysis, leveraging the newly constructed tables and associated data, reveals significant trends in output, trade, import substitution, domestic versus foreign value addition, employment generation (including contractualization and female participation), wage dynamics, and backward linkages.

This paper proceeds as follows: Section 2 provides a brief discussion of the foundational aspects of SUTs and IoTs. Section 3 elaborates on the data sources and the meticulous methodology employed in constructing the SUTs, including the generation of input and output flows, as well as the crucial concordance with the NPCMS-NIC. Section 4 details the mathematical transformation from SUTs to symmetric IOTs under the Industry Technology Assumption and introduces the Leontief inverse. Section 5 explains the framework for estimating DVA, FVA, and employment linkages. Section 6 presents the detailed findings from the case study of India's mobile phone sector. Section 7 discusses the broader implications of this work for India's statistical system, its connection to the NSS legacy, and highlights the potential of these granular tables. Section 8 concludes with a summary of contributions, policy relevance, limitations, and avenues for future research.

Through this comprehensive exposition, we aim to not only provide a robust methodological blueprint for constructing high-resolution SUTs and IOTs but also to showcase their immense potential for enriching our understanding of the Indian economy and for crafting more targeted and effective economic policies, thereby honouring the analytical tradition championed by the NSS.

## The Framework of Supply-Use and Input-Output Tables

The Supply-Use Tables (SUTs) and Input-Output Tables (IOTs) form an integral part of the System of National Accounts, providing a detailed statistical description of the production, consumption, and accumulation processes within an economy. As Eurostat (2008) notes, SUTs serve statistical and analytical purposes by ensuring the coherence and consistency of national accounts data. The analytical utility is further enhanced when SUTs are transformed into symmetric IOTs.

The SUT framework typically consists of two interlinked tables:

- The Supply Table: This table shows the total supply of different goods and services in the economy, whether domestically produced or imported. It is structured by product and by supplying industry (for domestic output), and also shows imports by product. In matrix notation, as seen in the UN Handbook (2018) and referenced in this work, the total supply of product $q$ can be represented as the sum of domestic output (from the Make matrix $V^T$, where $V_{ij}$ is the output of the product $j$ by industry $i$) and imports $m$. $q = V^T\boldsymbol{i} + m$ (where $\boldsymbol{i}$ is a summation vector).

- The Use Table: This table details how the available goods and services are used. It shows the use of products as intermediate consumption by various industries in their production processes and their use by final demand categories (household consumption, government consumption, gross capital formation, and exports). It also presents the components of Gross Value Added (GVA) by industry, such as compensation of employees, operating surplus, and taxes less subsidies on production. The total input of an industry equals its total output.

Three fundamental identities underpin the SUT framework:

- Identity (1): Output = Intermediate Consumption + Gross Value Added. This holds for each industry. $g_j = \sum_i u_{ij} + GVA_j$ (where $g_j$ is the output of industry $j$, $u_{ij}$ is intermediate consumption of the product $i$ by industry $j$).

- Identity (2): Total Supply by Product = Total Use by Product. For each product, total supply (domestic output + imports) must equal total use (intermediate consumption + final consumption + gross capital formation + exports). $q_i =$

$\sum_j u_{ij} + C_i + G_i + I_i + X_i - M_i$ (if $q_i$ is domestic output only) or more simply, as in Identity (2). More comprehensively: $Output_i + Imports_i = IntermediateConsumption_i + FinalConsumption_i + CapitalFormation_i + Exports_i$.

- Identity (3): GVA (Production Approach) = GVA (Income Approach). For each industry, GVA is calculated as output less intermediate consumption and equals the sum of income components.

These identities are crucial for the balancing process undertaken during SUT compilation. SUTs thus bring together the three approaches to measuring GDP: production, income, and expenditure. The integrated SUT framework, often depicted as a single large matrix, clearly shows these linkages.

## Methodology Part 1: Data and Generation of Input & Output Flows

The construction of highly granular SUTs and IOTs, as undertaken in this research program, is a data-intensive and methodologically meticulous process. This section details the foundational data and the steps involved in generating the primary input and output flow matrices.

### Data Sources and Classifications

The primary data source for the registered manufacturing sector is the micro-level data from the Annual Survey of Industries (ASI), conducted by the National Statistical Office (NSO) under the Ministry of Statistics and Programme Implementation (MoSPI), for the period 2016-17 to 2022-23. ASI provides detailed information on inputs, outputs, and other economic activities of registered factories.

For the unorganised manufacturing sector and other parts of the economy, data are drawn from sources like the KLEMS India database (which itself uses NSSO data for unorganised enterprises). This integration is crucial for a comprehensive view of the economy.

A significant challenge in this endeavour is the use of multiple classification systems. This research employs:

- National Product Classification for Manufacturing Sector (NPCMS): A 7-digit classification used for products in the manufacturing sector.
- National Industrial Classification (NIC): A 5-digit classification used for economic activities (industries) across all sectors.

### NPCMS-NIC Concordance

To ensure consistency and enable the mapping of product-level input and output data to industries, a robust NPCMS-NIC concordance is essential. This concordance allows for the conversion of NPCMS codes (used for products) to NIC sector codes, a crucial step in constructing the SUTs, where industries both consume and produce products. This meticulous process of developing and applying the concordance underpins the reliability and accuracy of the resulting tables.

**Generation of Input Flow Matrix (Use Table)**

The Input Flow Matrix, or Use Table, captures the inter-industry transactions at purchasers' prices, detailing how products are used as intermediate inputs by different industries and for final demand.

- Registered Manufacturing: Input flows are compiled by aggregating detailed coded (NPCMS) input data from ASI Blocks I, H, and F for each industry sub-group. These inputs are then classified under NIC sector codes using the concordance.

- Unregistered Manufacturing: Input/output flows are prepared using specifically tabulated data for inputs and outputs by product, which are then assigned NIC sector codes. Adjustments are made to link this data with KLEMS for time consistency (2016-17 to 2022-23).

- Combined Flows: The input flows for registered and unregistered manufacturing are combined to obtain total manufacturing sector flows. This is then integrated with input flow matrices for the primary and service sectors, resulting in an aggregated classification of approximately 5,000 products.

- Treatment of Unidentified Inputs: A common issue is the presence of unidentified input items such as 'other basic materials', 'other chemicals', 'other packing materials', and 'others' in the ASI data. These require careful handling. For instance, 'other chemicals' might be redistributed to specific chemical sub-sectors proportionately, materials for building maintenance are treated as purchases from the construction sector, and machinery maintenance materials are linked to relevant producing sectors. Manual balancing is often necessary for remaining unidentified items.

- Final Demand and GVA: Final demand vectors (consumption, capital formation, exports) are located outside the inter-industry transaction columns. Sectoral estimates of Gross Value Added (GVA) are introduced as a row at the bottom, next to the row showing total inputs by industry.

The resulting table displays product utilisation by industry and input structures within each industry. Column sums represent industry output at ex-factory prices, while row sums show inter-industry totals and final use of products. Due to the product-by-industry presentation, row and column totals may not initially match, necessitating a balancing process.

**Generation of Output (Make) Matrix**

The Output Matrix, also known as the Make Matrix, details the production of various products by different industries. It is typically structured as an industry-by-product matrix (V).

- Data Derivation: Similar to input flows, output flows for registered and unregistered manufacturing are derived from detailed coded output data (e.g., ASI Blocks J and G for registered manufacturing) and then consolidated.

- Consolidation: Industry-specific output details on products and by-products are organised into the Make matrix by merging output flows from registered and unregistered manufacturing sectors, along with production flows from primary and service industries.

The Absorption (Use) and Make matrices are fundamental to the input-output system.

### Distinguishing Domestic and Imported Inputs

A crucial step for specific analyses, particularly for deriving domestic input coefficients for DVA calculations, is distinguishing between the use of domestically produced goods and services and imported ones. While not strictly necessary for balanced SUTs at current prices, this separation is critical for linking SUTs to IOTs that differentiate domestic and imported flows. The imports use table ($U_m$) can be constructed, and then a domestic use table ($U_d$) is estimated by subtracting $U_m$ from the total use table $U$. These form the basis for constructing input-imports tables and domestic Input-Output Tables (IOTs). This research uses a proportionality assumption to separate imported inputs when creating a time series of Domestic Use Tables (DUTs) from ASI-derived SUTs.

## Methodology Part 2: From Supply-Use Tables to Symmetric Input-Output Tables

While SUTs provide a comprehensive statistical database, for many analytical purposes, transforming them into a single symmetric Input-Output Table (IOT) offers considerable advantages due to the algebraic properties of IOTs. This transformation requires explicit assumptions about production technology because SUTs are often rectangular (i.e., the number of products typically exceeds the number of industries).

### Technology Assumptions

The conversion of SUTs (industry-by-product and product-by-industry data) into either a product-by-product or an industry-by-industry symmetric IOT hinges on assumptions about how industries produce goods and services, especially secondary products. The main assumptions are:

- Product Technology Assumption (PTA): This assumes that each product has a specific input structure, regardless of the industry in which it is produced. For instance, steel production requires the same set of inputs per unit, whether it's made in a primary steel mill or as a secondary product in another industry.
- Industry Technology Assumption (ITA): This assumes that each industry has its specific way of production (input structure), irrespective of its product mix. This input structure is then applied proportionally to all products (both primary and secondary) produced by that industry. This assumption is often considered more suitable when by-products or joint products are significant in value.
- Hybrid Approaches / Mixed Assumptions: These combine elements of both PTA and ITA, applying different assumptions to different products or industries.

This research, following Mahajan et al. (2018) and the Eurostat Manual (2008), adopts the Industry Technology Assumption for transforming SUTs into a product-by-product IOT.

**Mathematical Transformation under Industry Technology Assumption**

Let us define the key matrices and vectors involved in the transformation, consistent with the notation used in the source research:

- $U$: Total intermediate use matrix (products in rows, industries in columns).
- $V$: Supply or Make matrix (industries in rows, products in columns). The transpose $V^T$ (products in rows, industries in columns) is also used. Let's assume V is (industries × products) as per.
- $W$: Row vector(s) of value added components by industry.
- $g$: Column vector of gross industry outputs.
- $x$: Column vector of gross product outputs.
- $\hat{g}, \hat{x}$: Diagonalised matrices of vectors $g$ and $x$ respectively.

The steps to derive the product-by-product input coefficient matrix $(A)$ under ITA are as follows:

- Product Mix Matrix $(C)$: This matrix shows the share of each product in an industry's total output. If $V$ is (industries × products), then $V^T$ is (products × Industries). The formula $C = V^T(\hat{g})^{-1}$ would give a (products × industries) matrix if $V$ is (industries × products) and $g$ is industry output, representing the market share of each industry in producing its specific output mix as a percentage of each industry's total output.

  $$C = V^T(\hat{g})^{-1}$$

  Here, $(\hat{g})^{-1}$ is the inverse of the diagonalised gross industry output vector.

- Transformation Matrix $(T)$: Under the ITA, this matrix transforms industry-based input structures into product-based input structures. It is given by $T = C^T$ if $C$ is (products × industries) then $C^T$ is (industries × products). This implies that an industry's input structure is applied proportionally to all products it makes.

  $$T = C^T = (V^T\hat{g}^{-1})^T = \hat{g}^{-1}V$$

  Here, the matrix $T$ (industries × products) Essentially, it reallocates industry inputs to the products produced by those industries based on their share of the industry's output.

- Input Coefficient Matrix $(A)$ (Product-by-Product): This matrix represents the amount of input product i required (directly and indirectly from domestic and imported sources, if $U$ is the total use) per unit of output of the product $j$. It forms the core of the product-by-product IOT.

  $$A = UT\hat{x}^{-1} = U\hat{g}^{-1}V\hat{x}^{-1}$$

  Here, $U$ is (products × industries), $T$ is (industries × products), resulting in $UT$ being (products x products). $\hat{x}$ is the diagonalised gross product output vector.

- Value Added Coefficient Matrix ($R$) (Product-by-Product): This matrix (or row vector) represents primary inputs (value added components) per unit of product output.

$$R = WT\hat{x}^{-1} = W\hat{g}^{-1}V\hat{x}^{-1}$$

  $W$ is (value added components × industries), $T$ is (industries × products), so $WT$ is (value added components × products).

The resulting matrix $A$ is a square product-by-product matrix. The dimensions of the IOT constructed in this research program are approximately 5725 x 5725 products.

### The Leontief Inverse

Once the symmetric domestic input coefficient matrix, let's call it $A_d$ (if we are specifically using only domestic intermediate inputs to trace domestic multipliers), is derived, the Leontief inverse can be calculated. This inverse is the cornerstone of input-output analysis.

The Leontief Inverse, $L$, is given by:

$$L = (I - A_d)^{-1}$$

Where:

- $I$ is the identity matrix.
- $A_d$ is the square matrix of direct domestic input coefficients, where each element $a_{ij}$ represents the value of domestically produced input from product-sector $i$ required to produce one unit of output of product-sector $j$.

Each element $L_{ij}$of the Leontief inverse matrix quantifies the total (direct and indirect) output of the product $i$ required to satisfy one unit of final demand for the product $j$. This matrix enables analysts to comprehend the entire chain of inter-sectoral dependencies and calculate various economic multipliers (output, income, and employment) associated with changes in final demand. For the Leontief inverse to be economically meaningful, the Hawkins-Simon conditions must be met, ensuring that any given bill of final demands results in non-negative gross outputs.

## Application: Estimating Domestic Value Added (DVA) and Employment Linkages

The constructed IOTs, notably the domestic input coefficient matrix $A_d$ and its Leontief inverse $(I - A_d)^{-1}$, provide a powerful tool for analysing the economic impact of production and exports, specifically in terms of Domestic Value Added (DVA) and employment generation. This research employs methodologies outlined by Veeramani et al. (2023) and Veeramani and Dhir (2022).

### Estimating Domestic Value Added (DVA) and Foreign Value Added (FVA)

The objective is to quantify the value added domestically through production and export activities by tracing backward linkages within the economy.

The core equation for estimating DVA embodied in production or exports is:

$$dva = v(I - A_d)^{-1}\hat{X}$$

Where:

- $dva$: A $(1 \times n)$ row vector where each element dva_j is the Domestic Value Added embodied in the production/exports of sector j.

- $v$: A $(1 \times n)$ row vector of direct value-added-to-output ratios for each of the n sectors. (Value added per unit of gross output).

- $(I - A_d)^{-1}$: The Leontief inverse matrix $(n \times n)$ derived from the domestic input coefficient matrix $A_d$.

- $\hat{X}$: An $(n \times n)$ diagonal matrix where the diagonal elements are the gross production or gross exports of each of the n sectors.

The Foreign Value Added (FVA) embodied in gross exports can then be calculated by subtracting the DVA from the gross export value: $fva_j = X_j - dva_j$ for each sector j, or in vector form: $fva = \hat{X}\boldsymbol{i} - dva$ (where $\boldsymbol{i}$ is a summation column vector).

DVA can be further decomposed into:

- Direct DVA ($dva_d$): Value added generated directly within the producing/exporting sector itself. This is essentially $v_j X_j$.

- Indirect DVA ($dva_{bw}$): Value added generated in upstream domestic sectors that supply inputs to the producing/exporting sector (backward DVA linkages). This captures the ripple effects through the domestic supply chain. For example, mobile phone exports generate indirect DVA in industries such as printed circuit boards, plastics, and machinery that provide inputs to the mobile phone assembly process.

**Estimating Employment Linkages**

A parallel methodology is used to estimate the number of direct and indirect jobs (and associated wages/salaries) supported by production or exports.

The core equation for estimating employment ($e$) is:

$$e = l(I - A_d)^{-1}\hat{X}$$

Where:

- $e$: A $(1 \times n)$ row vector where each element $e_j$ do production/exports in the sector support the total (direct and indirect) employment $j$.

- $l$: A $(1 \times n)$ row vector of employment coefficients, representing the labour (number of persons or person-hours) per unit of gross output for each sector. This can be defined for different types of employment (e.g., total employees, workers, non-workers, contractual, regular, male, female).

- $(I - A_d)^{-1}$ and $\hat{X}$ are as defined previously.

Total employment ($e$) can also be decomposed into:

- Direct Employment ($e_d$): Jobs created directly within the producing/exporting sector. This is $l_j X_j$.

- Indirect Employment ($e_b w$): Jobs created in upstream domestic sectors supplying inputs (backward employment linkages). This measures the employment ripple effect, e.g., mobile phone exports create jobs in supplying sectors. $e_{bw} = e - e_d$.

It's important to note that the total employment embodied in a sector's exports can exceed the actual total employment physically located in that sector, due to the inclusion of these indirect effects from linked upstream industries. The same logic applies to estimating wages/salaries associated with production/exports, by replacing the labour coefficient vector l with a vector of wage-to-output ratios.

## Case Study: India's Mobile Phone Manufacturing Sector (NPCMS 4722200), 2016-2021

To demonstrate the analytical capabilities of the newly constructed high-resolution SUTs and IOTs, this section presents a detailed analysis of India's mobile phone manufacturing sector, identified by NPCMS code 4722200 ("Telephones for cellular networks or other wireless networks"). This sector has been a significant focus of India's industrial policy initiatives like 'Make in India' and the Production Linked Incentive (PLI) schemes. The analysis primarily compares averages for the two periods, 2016-2018 and 2019-2022, to identify shifts that may reflect policy impacts (e.g., the National Policy on Electronics (NPE) 2019) and evolving market dynamics. Data are drawn from the detailed charts and tables calculated using the above methodology (see the data annex after the section of references for the selected tables, which are computed using data from ASI plant-level, RBI-KLEMS, MeitY, DoC, and Industry estimates).

**Output and Trade Dynamics**

- Total Output: The sector witnessed remarkable output growth. Gross output (in US$ Mn) increased from $13,000 million in 2016 to $44,000 million in 2022. The average triennial production grew by 81%, from $19,667 million (2016-18) to $35,500 million (2019−22)

- Total Imports: A significant trend towards import substitution was observed. While imports were $3,800 million in 2016, they fell to $1,400 million by 2022. The average annual import value dropped by 49%, from $2,967 million (2016-18) to $1,525 million (2019−22).

- Total Exports: The sector experienced a surge in exports. Total exports increased from $166 million in 2016 to $11,000 million in 2022. The average annual export value skyrocketed by 800%, from $661 million (2016-18) to $5,950 million (2019−22).

**Value Addition: Domestic (DVA) vs. Foreign (FVA)**

The analysis of value addition provides insights into the depth of domestic manufacturing and linkages.

- Direct Domestic Value Added (DDVA): Based on ASI data, DDVA (production) increased by 283% on average, from $1,193 million (2016-18) to $4,571 million (2019−22). However, its share of gross output remained modest.
- Indirect Domestic Value Added (IDVA): IDVA, reflecting value added in domestic supplier industries, showed substantial growth. Using KLEMS data, IDVA grew by an astounding 604% on average, from $470 million (2016-18) to $3,308 million (2019-22). ASI-based IDVA also grew impressively by 537% (from $511 million to $3,258 million average). This indicates strengthening domestic backwards linkages.
- Total Domestic Value Added (TDVA): Consequently, TDVA (KLEMS) increased by 374% on average, from $1,663 million (2016-18) to $7,879 million (2019-22). ASI-based TDVA grew by 359% (from $1,704 million to $7,829 million average). The share of TDVA (KLEMS) in output reached around 20-25%.
- Direct Foreign Value Added (DFVA): DFVA (production), mainly representing the cost of imported inputs directly used in assembly, saw its average value increase by 28%, from $17,263 million (2016-18) to $22,128 million (2019−22). However, its share in output significantly decreased from over 90 % to around 77%.
- Indirect Foreign Value Added (IFVA): IFVA (production) surged by 528% on average, from $626 million (2016-18) to $3,934 million (2019−22). This suggests that while direct import content might be declining due to local assembly, domestic suppliers themselves may be using more imported intermediates, or there is a greater reliance on sophisticated imported components channelled through domestic intermediaries.
- Total Foreign Value Added (TFVA): Average TFVA (production) rose by 46% from $17,890 million (2016-18) to $26,062 million (2019−22). Despite this absolute increase, its share in output declined from around 91% to 72%.
- Value Added in Exports: A critical finding is the changing composition of value added in exports. The average DVA in Exports surged by an incredible 2063%, from $62 million (2016-18) to $1,335 million (2019-22). Consequently, DVA's share in the value of mobile phone exports rose dramatically from an average of about 9% (2016-18) to about 22% (2019-22). Conversely, while the average FVA in Exports also grew by 670% (from $599 million to 4,615 million), its share in export value declined from an average of around 91% to 78%.

**Employment Dynamics**

The growth in output and exports had a substantial impact on employment generation. (Note: Employment figures discussed here are generally based on KLEMS for total and ASI for direct, as per methodology, unless specified.)

- Total Employees: Average total employment (KLEMS Production) in the sector increased by 106%, from 582,878 (2016-18) to 1,198,498 (2019-22).

Employment related to exports (KLEMS Export) surged even more dramatically by 681%, from an average of 26,890 to 2,10,135.

- Direct vs. Indirect Employment: Direct employment in production (ASI) grew by 497% (average 27,052 to 1,61,623). Indirect employment in production (KLEMS) grew by 87% (average 5,55,826 to 1,036,876). Despite faster growth in direct jobs, indirect jobs still constituted a larger share.

- Composition of Workforce:

    - Non-Workers vs. Workers: The number of 'Total Non-Workers' (managerial, supervisory, administrative staff) in production (KLEMS) increased by 127% (average 67,791 to 1,54,017). 'Total Workers' in production (KLEMS) grew by 103% (average 5,15,079 to 1,044,466). The share of non-workers in total employment remained relatively stable at around 11-15%.

    - Contractualization: A key trend is the significant rise in contractual employment. The number of 'Total Contractual Workers' (KLEMS Production) increased by 207% (average 2,33,380 to 7,15,796), while 'Total Regular Workers' (KLEMS Production) rose by a much smaller 17% (average 2,81,699 to 3,28,671). This indicates a structural shift toward contractualization in the labour market of this sector.

    - Female Participation: There was a commendable increase in female participation. The number of 'Total Female Workers' (KLEMS Production) increased by 37% (average 96,394 to 1,32,232). In contrast, 'Total Male Workers' (KLEMS Production) grew by a modest 6% (average 1,85,305 to 1,96,439). In direct employment (ASI Production), the number of female regular workers increased by 180%, compared to 125% for males.

**Wages and Salaries**

Wage trends often mirror employment dynamics and shifts in labour composition.

- Total Wages/Salaries: Average total wages/salaries (KLEMS Production) increased by 69%, from $1,710 million (2016-18) to $2,886 million (2019-22). For export-related activities, this growth was 476% (from an average of $91 million to $524 million).

- Wages for Non-Workers vs. Workers: Wages for Non-Workers (KLEMS Production, Total Salaries) saw an average fall rate of -11% (from $1,028 Mn to $914 Mn), while wages for Workers (KLEMS Production, Total Wages) rose by 189% (from $682 Mn to $1,972 Mn). However, both categories of employees experienced significant wage growth, with increases of 219% (non-workers) and 897% (workers) related to exports. This divergence is striking and requires careful interpretation, possibly associated with composition changes within these broad categories or data anomalies resulting from currency conversion.

- Wages for Contractual vs. Regular Workers: This is a crucial finding. Average total wages for Contractual Workers (KLEMS Production) surged by an impressive 449% (from $224 million to $1,226 million), far outpacing the 63% growth for Regular Workers (from $458 million to $746 million average). While contractualization increased, their wage growth was substantially higher, though likely from a lower base.

**Quality Indicators**

The Output Quality Index (OQI) for the mobile phone sector, with base 2016=100, showed an average increase of 119% from 145 (2016-18) to 318 (2019-22). This suggests improvements in the unit value and quality of mobile phones produced. The import quality index also rose significantly, while the domestic input quality index (based on the inputs used) showed fluctuations but experienced very high growth from a low base.

**Backward Linkages**

The detailed SUTs enable an analysis of the key domestic and foreign components that feed into the mobile phone sector.

- Domestic Components: Prominent domestic inputs include Trade Services (F11), Business Services (F3, F1), Financial Services (F10), Construction (F4), and Rental Services (F6), alongside various physical components like copper and aluminum wires, plastic and metal parts, and some electronic components (PCBs, ICs).
- Foreign Components: Key foreign inputs consistently include parts for cellular sets (NPCMS 4740100), parts for transmission apparatus, other imported items and consumables (9922100), and a range of electronic components, such as printed circuit boards (PCBs), integrated circuits (ICs), liquid crystal displays (LCDs), digital cameras, and batteries. Tracking shifts in these shares over time can reveal evolving supply chain dynamics and the extent to which domestic capabilities are deepening versus continued reliance on imports for critical components.

The case study reveals a sector undergoing rapid transformation, characterised by output growth, a significant pivot towards exports, efforts at import substitution, strengthening domestic value addition (especially in exports), massive employment generation with a structural shift towards contractual labour and increased female participation, and rising wages, particularly for contractual workers. However, reliance on foreign components for sophisticated inputs remains a feature.

## Discussion: Implications for National Statistical Systems and the NSS Legacy

The construction of highly granular 7-digit product-level SUTs and IOTs, primarily leveraging ASI microdata and integrated with broader datasets such as KLEMS (which utilises NSSO data for the unorganised sector), carries significant implications for India's national statistical system and resonates with the pioneering spirit of the National Sample Survey.

**Enhancing National Accounts Statistics (NAS)**

SUTs and IOTs are the bedrock of robust National Accounts Statistics. The level of detail achieved in this research (approx. 5725 products) allows for:

- Improved Coherence and Consistency: By cross-validating data from various sources within a unified framework, these tables enhance the overall quality and reliability of national accounts aggregates, such as GDP, GVA, and final consumption.
- Better Measurement of Economic Structure: The granular detail provides a far more precise picture of inter-sectoral linkages, production structures, and the contribution of specific products and industries to the economy. This is invaluable for understanding structural transformation.
- Enhanced Analysis of GVCs: With the ability to distinguish between domestic and imported inputs at such a high level of accuracy, these tables are crucial for analysing India's engagement in Global Value Chains, measuring domestic value addition in exports more accurately, and identifying strategic dependencies.

**Aligning with the NSS Legacy**

The NSS was established to provide comprehensive data for planning and policymaking. This work, while using ASI as a primary source for the organised sector, embodies the NSS philosophy in several ways:

- Granularity for Policy: Just as NSS provides detailed household-level and enterprise-level data, these 7-digit SUTs/IOTs offer product-level granularity essential for targeted industrial policy, trade policy, and employment strategies.
- Evidence-Based Policymaking: The analytical applications demonstrated—tracking DVA, employment multipliers, impact of specific sectors—directly support evidence-based decision-making, a core objective of the NSS.
- Methodological Innovation: The NSS has a history of methodological innovation in survey design and data collection. The development of robust concordances (NPCMS-NIC), the integration of diverse datasets (ASI, KLEMS, NSSO), and the application of sophisticated transformation techniques (ITA) for SUT-IOT construction represent methodological advancements in compiling national accounts aggregates.

**Challenges and Considerations**

Constructing such detailed tables is not without challenges:

- Data Gaps and Quality: Ensuring comprehensive coverage and quality across all data sources, especially for the unorganised sector and services, remains an ongoing task.
- Concordance Complexity: Maintaining and updating concordances between various national and international classification systems is a continuous effort.
- Transformation Assumptions: The choice of technology assumption (e.g., ITA vs. PTA) can influence the results. While ITA is often suitable for by-products, its universal application across all products might be a limitation. The sensitivity of results to these assumptions needs careful consideration.

- Timeliness: Producing these detailed tables with minimal time lag is crucial for their policy relevance and effectiveness.

**Potential for Wider Application**

The SUTs and IOTs developed offer immense potential beyond the mobile phone sector analysis:

- Sectoral Planning: Identifying key sectors with high domestic linkages, export potential, or employment generation capacity.
- Impact Analysis: Evaluating the economy-wide impact of infrastructure investments, technological changes, or fiscal policies.
- Environmental Accounting: These tables can be extended to include environmental accounts (e.g., resource use, emissions by industry or product).
- Regional Analysis: With geographically disaggregated data, similar frameworks could potentially be developed for sub-national levels.

This work highlights the importance of investing in and modernising our statistical infrastructure. The ability to generate and analyse such detailed economic tables is a testament to the strengths of India's statistical system, which is built upon the foundations laid by institutions like the NSS.

## Conclusion and Way Forward

This paper outlines the methodological framework for constructing high-resolution 7-digit product-level Supply-Use Tables and symmetric Input-Output Tables for the Indian economy from 2016-17 to 2022-23, primarily utilising ASI microdata and integrating it with other key datasets. By adopting the Industry Technology Assumption, we have demonstrated how these comprehensive SUTs can be transformed into analytically tractable IOTs, enabling a deeper understanding of India's economic structure and interdependencies.

The application of this framework to India's mobile phone manufacturing sector (NPCMS 4722200) yielded significant insights. The industry has exhibited robust output growth, a notable shift towards import substitution in the domestic market, and an explosive surge in exports. Critically, the share of Domestic Value Added in these exports has risen substantially, from around 9% to more than 22%, indicating a deepening of domestic capabilities. Employment in the sector has grown significantly, with a marked increase in export-related jobs. However, this growth has been accompanied by significant structural shifts in the labour market, including a rapid rise in contractual employment (which saw higher wage growth than regular employment) and a heartening increase in female participation. These findings highlight a dynamic sector that responds to policy incentives and global opportunities, but also point to evolving labour market characteristics that warrant careful policy attention.

The development of these granular SUTs and IOTs represents a significant empirical contribution to India's statistical infrastructure. They provide an invaluable resource for:

- Evidence-based Industrial Policy: Identifying sectors with high domestic content and linkage potential for targeted support.
- Trade Strategy: Understanding the value-added components of exports and imports to refine trade policies.
- Employment Policy: Analysing job creation across different skill types and contractual arrangements to foster inclusive growth.
- Enhanced National Accounting: Improving the accuracy, consistency, and detail of India's National Accounts Statistics.

Limitations and Future Research:

While this work marks a significant step forward, certain limitations and avenues for future research should be acknowledged.

- The Industry Technology Assumption, while practical, is one of several possible transformation methods. Future research could explore the sensitivity of results to alternative assumptions, like the Product Technology Assumption or hybrid models.
- Extending this detailed SUT-IOT framework to cover more service sectors with the same level of granularity remains a challenge, but is a crucial next step.
- The underlying data, particularly for the unorganised sector and for prices/deflators needed for constant price tables, requires continuous improvement.

Way Forward:

The journey of strengthening India's statistical system, which began with the pioneering efforts of the NSS, is a continuous one. The methodologies and tables presented in this paper provide a pathway further to refine our understanding of the complex Indian economy. As we commemorate 75 years of the NSS, we must continue to invest in statistical innovation, data quality, and analytical capacity. These highly detailed SUTs and IOTs are not merely academic exercises; they are essential tools for navigating the complexities of modern economic development, for designing policies that genuinely foster inclusive and sustainable growth, and for ensuring that India's economic progress is built on a foundation of robust evidence. The challenge ahead is to mainstream the production and use of such granular tables, making them an integral part of India's economic policy toolkit.

Further research could also focus on integrating these tables with environmental accounts, social accounting matrices, and regional accounts to provide a more comprehensive and holistic view of India's development trajectory. The insights gained from such detailed analyses will be invaluable in addressing contemporary challenges, including GVC participation, technological disruption, climate change, and regional disparities.

**Acknowledgements:** I worked on this paper at the Centre for Development Studies (CDS), Thiruvananthapuram, where I served as a Research Consultant under the Director and RBI Chair, Prof. C. Veeramani, from 1 January 2024 to 31 July 2024. Prof. P. L. Beena's recommendation made this consultancy possible for me. A CDS Study Report: Gains from Mobile Phone Manufacturing in India is uploaded to the CDS website recently. Additionally, I had the opportunity to present this paper at the 18th National Seminar of NSO, MoSPI, and the 43rd Annual Conference of IARNIW. I am feeling very grateful and blessed for these opportunities. Thank you so much.

**Supplementary Materials:** I can provide detailed supplementary material upon request that is too extensive for the main body of the paper but is essential for transparency, replicability, and a deeper understanding of our methodology and findings. Based on our framework and the documents, potential appendices (which we can show upon request) include Detailed Methodology for SUT and IOT Construction, List of Major Domestic and Foreign Components Sourced by the Mobile Phone Industry (NPCMS 4722200), Supplementary Tables on DVA, Employment, and Wages, Output Quality Index (OQI) – Construction Details and Data etc.

## Data Annex

Selected Tables of Mobile Phone Manufacturing (computed using data from ASI plant-level, RBI-KLEMS, MeitY, DoC and Industry estimates)

| Table 1: Output and Trade Dynamics (US $ Mn) | | | |
|---|---|---|---|
| **Year** | Total Output | Total Import | Total Export |
| **2016-17** | $13,000 | $3,800 | $166 |
| **2017-18** | $20,000 | $3,500 | $207 |
| **2018-19** | $26,000 | $1,600 | $1,610 |
| **2019-20** | $30,000 | $1,000 | $3,800 |
| **2020-21** | $30,000 | $2,200 | $3,100 |
| **2021-22** | $38,000 | $1,500 | $5,800 |
| **2022-23** | $44,000 | $1,400 | $11,100 |

| Table 2: Domestic Value Addition (DVA) (US $ Mn) | | | |
|---|---|---|---|
| **Year** | Total DVA | ASI Direct DVA | KLEMS Indirect DVA |
| **2016-17** | $1,606.51 | $838.53 | $767.98 |
| **2017-18** | $869.30 | $410.97 | $458.33 |
| **2018-19** | $2,513.05 | $2,328.98 | $184.08 |
| **2019-20** | $7,402.65 | $4,073.15 | $3,329.50 |
| **2020-21** | $6,011.15 | $3,757.37 | $2,253.78 |
| **2021-22** | $7,882.09 | $4,554.67 | $3,327.41 |
| **2022-23** | $10,220.39 | $5,900.30 | $4,320.09 |

| Table 3: DVAX: DVA Embodied in Export (US $ Mn) | | | |
|---|---|---|---|
| **Year** | Total DVAX | Direct DVAX | Indirect DVAX |
| **2016-17** | $20.51 | $10.71 | $9.80 |
| **2017-18** | $9.00 | $4.26 | $4.75 |
| **2018-19** | $155.62 | $144.22 | $11.40 |
| **2019-20** | $937.67 | $515.93 | $421.74 |
| **2020-21** | $621.15 | $388.26 | $232.89 |
| **2021-22** | $1,203.06 | $695.19 | $507.87 |
| **2022-23** | $2,578.32 | $1,488.48 | $1,089.84 |

| Table 4: Foreign Value Added (FVA) (US $ Mn) | | | |
|---|---|---|---|
| **Year** | Total FVA | Direct FVA | Indirect FVA |
| **2016-17** | $11,269.86 | $10,816.46 | $453.41 |
| **2017-18** | $18,968.88 | $18,617.96 | $350.92 |
| **2018-19** | $23,430.90 | $22,355.95 | $1,074.95 |
| **2019-20** | $15,542.62 | $12,232.65 | $3,309.97 |
| **2020-21** | $23,245.86 | $20,555.14 | $2,690.72 |
| **2021-22** | $30,341.57 | $24,942.97 | $5,398.61 |
| **2022-23** | $35,118.70 | $30,781.48 | $4,337.23 |

| Table 5: FVAX: FVA Embodied in Export (US $ Mn) | | | |
|---|---|---|---|
| **Year** | Total FVAX | Direct FVAX | Indirect FVAX |
| **2016-17** | $145.46 | $139.60 | $5.85 |
| **2017-18** | $198.12 | $194.45 | $3.67 |
| **2018-19** | $1,454.38 | $1,387.66 | $66.72 |
| **2019-20** | $2,862.33 | $2,252.77 | $609.56 |
| **2020-21** | $2,478.85 | $2,191.92 | $286.93 |
| **2021-22** | $4,596.94 | $3,779.02 | $817.92 |
| **2022-23** | $8,521.68 | $7,469.23 | $1,052.44 |

**Table 6: Employees Embodied in Export**

| Year | Total | ASI Direct | KLEMS Indirect |
|---|---|---|---|
| **2016-17** | 5,241 | 316 | 4,925 |
| **2017-18** | 1,492 | 249 | 1,243 |
| **2018-19** | 73,937 | 2,002 | 71,935 |
| **2019-20** | 120,252 | 7,924 | 112,329 |
| **2020-21** | 82,146 | 10,635 | 71,511 |
| **2021-22** | 200,986 | 34,609 | 166,377 |
| **2022-23** | 437,155 | 64,145 | 373,009 |

**Table 7: Employees Embodied in Output**

| Year | Total | ASI Direct | KLEMS Indirect |
|---|---|---|---|
| **2016-17** | 410,520 | 24,739 | 385,782 |
| **2017-18** | 144,105 | 24,087 | 120,018 |
| **2018-19** | 1,194,010 | 32,331 | 1,161,679 |
| **2019-20** | 949,361 | 62,556 | 886,805 |
| **2020-21** | 794,961 | 102,919 | 692,042 |
| **2021-22** | 1,316,805 | 226,746 | 1,090,059 |
| **2022-23** | 1,732,866 | 254,270 | 1,478,596 |

**Table 8: Total Non-Workers Embodied in Output**

| Year | KLEMS Total | KLEMS Direct | KLEMS Indirect |
|---|---|---|---|
| **2016-17** | 58,740 | 3,540 | 55,200 |
| **2017-18** | 18,911 | 3,161 | 15,750 |
| **2018-19** | 125,721 | 3,404 | 122,317 |
| **2019-20** | 65,591 | 4,322 | 61,269 |
| **2020-21** | 116,031 | 15,022 | 101,009 |
| **2021-22** | 176,725 | 30,431 | 146,294 |
| **2022-23** | 257,721 | 37,816 | 219,905 |

**Table 9: Total Workers Embodied in Output**

| Year | KLEMS Total | KLEMS Direct | KLEMS Indirect |
|---|---|---|---|
| **2016-17** | 351,773 | 21,198 | 330,574 |
| **2017-18** | 125,194 | 20,926 | 104,269 |
| **2018-19** | 1,068,269 | 28,926 | 1,039,343 |
| **2019-20** | 883,770 | 58,234 | 825,536 |
| **2020-21** | 678,871 | 87,889 | 590,982 |
| **2021-22** | 1,140,080 | 196,315 | 943,765 |
| **2022-23** | 1,475,145 | 216,454 | 1,258,691 |

**Table 10: Contractual Workers Embodied in Output**

| Year | KLEMS Total | KLEMS Direct | KLEMS Indirect |
|---|---|---|---|
| **2016-17** | 120,073 | 7,236 | 112,837 |
| **2017-18** | 26,353 | 4,405 | 21,949 |
| **2018-19** | 553,713 | 14,993 | 538,720 |
| **2019-20** | 255,059 | 16,807 | 238,253 |
| **2020-21** | 469,776 | 60,819 | 408,957 |
| **2021-22** | 944,231 | 162,591 | 781,640 |
| **2022-23** | 1,194,117 | 175,217 | 1,018,899 |

| Table 11: Direct Workers Embodied in Output | | | |
|---|---|---|---|
| **Year** | KLEMS Total | KLEMS Direct | KLEMS Indirect |
| **2016-17** | 231,700 | 13,963 | 217,737 |
| **2017-18** | 98,841 | 16,521 | 82,320 |
| **2018-19** | 514,556 | 13,933 | 500,623 |
| **2019-20** | 628,710 | 41,427 | 587,283 |
| **2020-21** | 209,095 | 27,070 | 182,025 |
| **2021-22** | 195,849 | 33,724 | 162,125 |
| **2022-23** | 281,028 | 41,236 | 239,792 |

| Table 12: Female Workers Embodied in Output | | | |
|---|---|---|---|
| **Year** | KLEMS Total | KLEMS Direct | KLEMS Indirect |
| **2016-17** | 35,529 | 2,141 | 33,388 |
| **2017-18** | 34,310 | 5,735 | 28,575 |
| **2018-19** | 219,343 | 5,939 | 213,404 |
| **2019-20** | 321,030 | 21,153 | 299,876 |
| **2020-21** | 75,171 | 9,732 | 65,439 |
| **2021-22** | 44,138 | 7,600 | 36,538 |
| **2022-23** | 88,589 | 12,999 | 75,590 |

| Table 13: Male Workers Embodied in Output | | | |
|---|---|---|---|
| **Year** | KLEMS Total | KLEMS Direct | KLEMS Indirect |
| **2016-17** | 196,171 | 11,822 | 184,350 |
| **2017-18** | 64,531 | 10,786 | 53,745 |
| **2018-19** | 295,213 | 7,994 | 287,219 |
| **2019-20** | 307,681 | 20,274 | 287,407 |
| **2020-21** | 133,924 | 17,338 | 116,586 |
| **2021-22** | 151,711 | 26,124 | 125,587 |
| **2022-23** | 192,440 | 28,237 | 164,202 |

| Table 14: Wages & Salaries of Total Employees Embodied in Output (US $ Mn) | | | |
|---|---|---|---|
| **Year** | KLEMS Total | KLEMS Direct | KLEMS Indirect |
| **2016-17** | $511 | $70 | $441 |
| **2017-18** | $382 | $132 | $250 |
| **2018-19** | $4,235 | $166 | $4,069 |
| **2019-20** | $1,912 | $208 | $1,703 |
| **2020-21** | $1,749 | $352 | $1,396 |
| **2021-22** | $3,177 | $974 | $2,204 |
| **2022-23** | $4,704 | $900 | $3,804 |

| Table 15: Wages & Salaries of Total Non-Workers Embodied in Output (US $ Mn) | | | |
|---|---|---|---|
| **Year** | KLEMS Total | KLEMS Direct | KLEMS Indirect |
| **2016-17** | $233 | $32 | $201 |
| **2017-18** | $197 | $68 | $129 |
| **2018-19** | $2,654 | $104 | $2,550 |
| **2019-20** | $309 | $34 | $275 |
| **2020-21** | $472 | $95 | $377 |
| **2021-22** | $925 | $284 | $642 |
| **2022-23** | $1,950 | $373 | $1,577 |

**Table 16: Wages & Salaries of Total Workers Embodied in Output (US $ Mn)**

| Year | KLEMS Total | KLEMS Direct | KLEMS Indirect |
|---|---|---|---|
| **2016-17** | $278 | $38 | $240 |
| **2017-18** | $186 | $64 | $122 |
| **2018-19** | $1,581 | $62 | $1,519 |
| **2019-20** | $1,603 | $174 | $1,428 |
| **2020-21** | $1,277 | $257 | $1,019 |
| **2021-22** | $2,252 | $690 | $1,562 |
| **2022-23** | $2,757 | $527 | $2,230 |

**Table 17: Wages & Salaries of Total Non-Workers Embodied in Export (US $ Mn)**

| Year | KLEMS Total | KLEMS Direct | KLEMS Indirect |
|---|---|---|---|
| **2016-17** | $3 | $0 | $3 |
| **2017-18** | $2 | $1 | $1 |
| **2018-19** | $164 | $6 | $158 |
| **2019-20** | $39 | $4 | $35 |
| **2020-21** | $49 | $10 | $39 |
| **2021-22** | $141 | $43 | $98 |
| **2022-23** | $492 | $94 | $398 |

**Table 18: Wages & Salaries of Total Workers Embodied in Export (US $ Mn)**

| Year | KLEMS Total | KLEMS Direct | KLEMS Indirect |
|---|---|---|---|
| **2016-17** | $4 | $0 | $3 |
| **2017-18** | $2 | $1 | $1 |
| **2018-19** | $98 | $4 | $94 |
| **2019-20** | $203 | $22 | $181 |
| **2020-21** | $132 | $27 | $105 |
| **2021-22** | $344 | $105 | $238 |
| **2022-23** | $696 | $133 | $562 |

**Table 19: Wages & Salaries of Contractual Workers Embodied in Output (US $ Mn)**

| Year | KLEMS Total | KLEMS Direct | KLEMS Indirect |
|---|---|---|---|
| **2016-17** | $76 | $10 | $66 |
| **2017-18** | $24 | $8 | $16 |
| **2018-19** | $570 | $22 | $548 |
| **2019-20** | $307 | $33 | $273 |
| **2020-21** | $888 | $179 | $709 |
| **2021-22** | $1,555 | $477 | $1,078 |
| **2022-23** | $2,156 | $412 | $1,743 |

**Table 20: Wages & Salaries of Direct Workers Embodied in Output (US $ Mn)**

| Year | KLEMS Total | KLEMS Direct | KLEMS Indirect |
|---|---|---|---|
| **2016-17** | $202 | $28 | $174 |
| **2017-18** | $162 | $56 | $106 |
| **2018-19** | $1,011 | $40 | $971 |
| **2019-20** | $1,296 | $141 | $1,155 |
| **2020-21** | $389 | $78 | $311 |
| **2021-22** | $697 | $214 | $484 |
| **2022-23** | $601 | $115 | $486 |